\documentclass[preprint,12pt]{elsarticle} 
\usepackage{graphicx} 
\usepackage{amssymb} 
\usepackage{amsmath} 
\usepackage{amsfonts} 
\usepackage{slashed} 
\usepackage[dvipsnames]{xcolor} 
\newcommand{\be}{\begin{equation}} 
\newcommand{\ee}{\end{equation}}

\journal{Physics Letters B} 

\begin{document} 

\begin{frontmatter}

\title{Compatibility of recent $\Xi$-nuclear bound state signals} 

\author[a]{E.~Friedman} 
\author[a]{A.~Gal\corref{cor2}} 
\address[a]{Racah Institute of Physics, The Hebrew University, 9190400 
Jerusalem, Israel} 
\cortext[cor2]{corresponding author: Avraham Gal, avragal@savion.huji.ac.il}  

\begin{abstract} 

J-PARC E05 reported recently a hint of a $\Xi^--{^{11}{\rm B}}$ nuclear 
state in the ${^{12}}{\rm C}(K^-,K^+){^{12}_{\Xi}}{\rm Be}$ % end-point 
spectrum, bound by $B_{\Xi^-}=8.9\pm 1.4 {^{+3.8}_{-3.1}}$ MeV. Using 
a density-dependent $\Xi$-nuclear optical potential $V_{\rm opt}^{\Xi}(\rho)$ 
we explore to what extent a $\Xi^-_{1s}$ assignment of this nuclear state is 
compatible with $\Xi^-_{1s}$ and $\Xi^-_{1p}$ nuclear-state interpretations 
of $\Xi^-$ capture events in light emulsion-nuclei experiments. We find that 
the only acceptable $\Xi^-_{1s}$ assignment at present, barring an abnormally 
strong repulsive $\rho^2$ component of $V_{\rm opt}^{\Xi}(\rho)$, is that for 
the $\Xi^--{^{11}{\rm B}}$ signal. This finding supports reassigning $\Xi^-$ 
capture events in $^{14}$N, originally assigned as $\Xi^-_{1s}-{^{14}{\rm N}}$ 
nuclear states, to $\Xi^0_{1p}-{^{14}{\rm C}}$ nuclear states. The depth of 
$V_{\rm opt}^{\Xi}(\rho)$ at nuclear-matter density $\rho_0=0.17$~fm$^{-3}$ 
is then $-V_{\rm opt}^{\Xi}(\rho_0)\approx 21$~MeV. 

\end{abstract} 

\begin{keyword}
Hyperon strong interaction. Cascade hypernuclei. Optical model fits.  
\end{keyword} 

\end{frontmatter}

\section{Introduction}
\label{sec:intro}

Little is known experimentally about the doubly strange (${\cal S}=-2$) 
$\Lambda\Lambda$ and $\Xi N$ two-baryon systems~\cite{GHM16,HN18}. 
The ${\cal S}=-2$ dynamics is likely to play a decisive role in dense 
neutron-star matter, see Ref.~\cite{Vidana25} for a very recent discussion. 
Regarding the $\Xi N$ interaction, a recent measurement of $\Xi^-p$ 
correlations in $p$-Pb ultra-relativistic collisions at the 
LHC~\cite{ALICE19,ALICE20} established that the low-energy $\Xi^-p$ 
interaction is attractive, in general agreement with HAL-QCD lattice QCD 
calculations~\cite{HALQCD20}. No similar experimental verification is 
available yet for other charge states of the two-body $\Xi N$ system. However, 
complementary information may be derived by studying $\Xi$-nuclear systems, 
which is the subject of the present work. 

\begin{figure}[!ht] 
\begin{center} 
\includegraphics[width=0.7\textwidth]{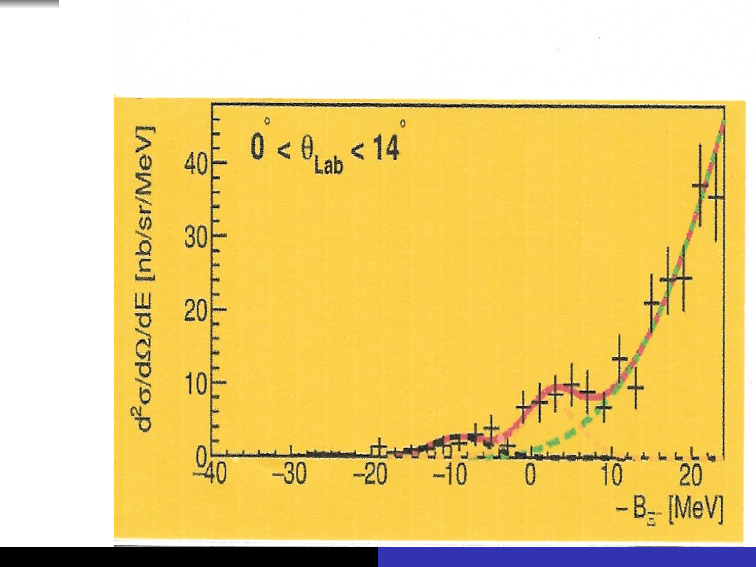} 
\caption{Best fit of J-PARC E05 $^{12}$C($K^-,K^+$) endpoint 
spectrum~\cite{E05}. Enhancements at $B_{\Xi^-}=8.9\pm 1.4 {^{+3.8}_{-3.1}}$ 
and $B_{\Xi^-}=-2.4\pm 1.3{^{+2.8}_{-1.2}}$~MeV are marked in red, see text.} 
\label{fig:12C} 
\end{center} 
\end{figure} 

Very recently the J-PARC E05 experiment~\cite{E05} reported a measurement of 
the $^{12}$C($K^-,K^+$) inclusive reaction spectrum at $p_{K^-}$=1.8~GeV/c, 
with energy resolution of 8.2~MeV (FWHM). Enhancements below and near the 
${^{12}}{\rm C}(K^-,K^+){^{12}_{\Xi}}{\rm Be}$ continuum threshold were 
observed, see Fig.~\ref{fig:12C}. A hint of $\Xi^--{^{11}{\rm B}}$ bound 
state at $B_{\Xi^-}=8.9\pm 1.4 {^{+3.8}_{-3.1}}$~MeV was suggested for the 
lower enhancement, most likely a $\Xi^-_{1s}-{^{11}{\rm B}}$ $J^{\pi}=1^-$ 
state~\cite{DG83}. Previous, poorer resolution $^{12}$C($K^-,K^+$) 
experiments at KEK~\cite{E224} and BNL~\cite{E885,E906} did not observe 
it~\cite{GHM16,HN18}. The upper enhancement, at $B_{\Xi^-}=-2.4\pm 1.3 
{^{+2.8}_{-1.2}}$~MeV, would then belong to the $\Xi^-_{1p}-{^{11}{\rm B}}$ 
configuration, superposed on a sharply rising quasi-free excitation spectrum. 
It is dominated by a $J^{\pi}=2^+$ state with a production cross section 
about 7 times larger than that of the $J^{\pi}=1^-$ member of the 
$\Xi^-_{1s}-{^{11}{\rm B}}$ ground-state configuration~\cite{DG83}. 

In this work we explore to what extent a $\Xi^-_{1s}$ assignment for the 
recent hint of a $\Xi^- - {^{11}{\rm B}}$ nuclear state in the J-PARC E05 
${^{12}}{\rm C}(K^-,K^+){^{12}_{\Xi}}{\rm Be}$ end-point spectrum~\cite{E05} 
is compatible with $\Xi^-_{1s}$ and $\Xi^-_{1p}$ nuclear-state interpretations 
of $\Xi^-$ capture events recorded in light emulsion-nuclei~\cite{E176,KISO,
IBUKI,E07}. To this end we use a density-dependent optical-potential 
$V_{\rm opt}^Y(\rho)$ ($Y$=hyperon) applied by us recently~\cite{FG23b,FG23c} 
to $\Lambda$ hypernuclear states observed in several types of production 
experiments across the periodic table, as well as in a limited form to $\Xi^-$ 
hypernuclei~\cite{FG21}. In a subsequent work~\cite{FG23a} we questioned 
$\Xi^-_{1s}-{^{14}{\rm N}}$ assignments of KEK-E373 and J-PARC E07 capture 
events in $^{14}$N made in Ref.~\cite{E07}. Here we find that the only 
acceptable $\Xi^-_{1s}$ assignment at present, barring an abnormally strong 
repulsive $\rho^2$ component of $V_{\rm opt}^{\Xi}(\rho)$, is that for the 
J-PARC E05 $\Xi^--{^{11}{\rm B}}$ bound-state signal. 
This finding supports reassigning $\Xi^-$ capture events in $^{14}$N, 
originally assigned as $\Xi^-_{1s}-{^{14}{\rm N}}$ nuclear states~\cite{E07}, 
to $\Xi^0_{1p}-{^{14}{\rm C}}$ nuclear states~\cite{FG23a}. The depth of 
$V_{\rm opt}^{\Xi}(\rho)$ at nuclear-matter density $\rho_0=0.17$~fm$^{-3}$ 
is then $-V_{\rm opt}^{\Xi}(\rho_0)\approx 21$~MeV. Other theoretical 
works~\cite{Sun16,Tanimura22,Isaka24} using various many-body methodologies 
(RMF, SHF, AMD) while accepting one of the few reported binding-energy values 
for $\Xi^-_{1s}-{^{14}{\rm N}}$~\cite{KISO,E07} find a smaller $\Xi$-nucleus 
potential depth of about 14~MeV, thereby predicting binding energy of only 
a few MeV for $\Xi^-_{1s}-{^{11}{\rm B}}$. 

Below we proceed as follows. In Sect.~\ref{sec:method} we specify the 
density-dependent $\Xi$-nuclear optical potential $V_{\rm opt}^{\Xi}(\rho)$ 
used in the present work to calculate $\Xi^-$ nuclear binding energies. 
In Sect.~\ref{sec:capture} we list and discuss briefly the $\Xi^-$ capture 
data available in $^{12}$C and $^{14}$N, some of which are used alongside with 
the $\Xi^--{^{11}{\rm B}}$ hint of a nuclear bound state reported recently by 
J-PARC E05~\cite{E05}. Results of $\Xi^-$ nuclear binding-energy calculations 
using $V_{\rm opt}^{\Xi}(\rho)$ are reported in Sect.~\ref{sec:calc} and 
concluding remarks are made in Sect.~\ref{sec:concl}.

\section{Optical potential methodology}
\label{sec:method}

$\Xi^-$ atomic and nuclear bound states in $N=Z$ nuclei such as $^{12}$C and
$^{14}$N are calculated using a finite-size Coulomb potential $V_c^{\Xi^-}$,
including vacuum-polarization terms, plus a $\Xi$-nuclear density-dependent 
optical potential, 
\begin{equation} 
V_{\rm opt}^{\Xi}(\rho)=V_{\Xi}^{(2)}(\rho)+V_{\Xi}^{(3)}(\rho) 
\label{eq:V}, 
\end{equation} 
consisting of terms representing two-body $\Xi N$ and three-body 
$\Xi NN$ interactions: 
\begin{equation} 
V_{\Xi}^{(2)}(\rho) = -\frac{4\pi}{2\mu_{\Xi}}f^{(2)}_A\,b_0^A(\rho)\,\rho, 
\label{eq:V2} 
\end{equation}  
\begin{equation} 
V_{\Xi}^{(3)}(\rho) = +\frac{4\pi}{2\mu_{\Xi}}f^{(3)}_A\,B_0\,
\frac{\rho^2}{\rho_0}, 
\label{eq:V3} 
\end{equation} 
with strength parameters $b_0^A(\rho),B_0$ in units of fm ($\hbar=c=1$). 
In these expressions $A$ is the mass number of the {\it nuclear core}, $\rho$ 
is a nuclear density normalized to $A$, $\mu_{\Xi}$ is the $\Xi^-$-nucleus 
reduced mass and $f^{(2,3)}_A$ are kinematical factors transforming 
$b_0^A(\rho)$ and $B_0$ from the $\Xi N$ and $\Xi NN$ 
center-of-mass (c.m.) systems, respectively, to the $\Xi$-nucleus c.m. system: 
\begin{equation} 
f^{(2)}_A=1+\frac{A-1}{A}\frac{\mu_{\Xi}}{m_N},\,\,\,\,\,\,
f^{(3)}_A=1+\frac{A-2}{A}\frac{\mu_{\Xi}}{2m_N}. 
\label{eq:fA} 
\end{equation} 
The effective density-dependent $\Xi N$ isoscalar c.m. scattering amplitude 
$b_0^A(\rho)$ is given by 
\begin{equation} 
b_0^A(\rho) = \frac{{\rm Re}\,b_0}{1+(3k_F/2\pi)f^{(2)}_A{\rm Re}\,b_0} + 
{\rm i}\,{\rm Im}\,b_0,
\label{eq:b0} 
\end{equation} 
where $k_F$ is the Fermi momentum associated with density $\rho$: 
$k_F=(3{\pi}^2\rho/2)^{1/3}$. The density dependence of $b_0^A(\rho)$ accounts 
for long-range Pauli correlations in $\Xi N$ in-medium multiple scatterings, 
starting at $\rho^{4/3}$ in a nuclear-density expansion~\cite{DHL71,WRW97} 
as practised in our $K^-$-atom studies~\cite{FG17}. Introducing it also in 
$V_{\Xi}^{(3)}$ makes little difference, which is why it is skipped in 
Eq.~(\ref{eq:V3}). Short-range $NN$ correlations, starting at order $\rho^2$ 
of the optical-potential density expansion, have little effect at densities 
below $\rho_0$ where the main contribution arises in the present context from 
three-body $\Xi NN$ interactions. Precisely the same functional form of 
$V_{\rm opt}(\rho)$, including $V^{(3)}(\rho)$, was used by us recently for 
constructing the $\Lambda$-nuclear optical potential $V_{\rm opt}^{\Lambda}
(\rho)$~\cite{FG23b,FG23c}. Three-body $YNN$ interactions appear at order 
N$^2$LO in EFT formulations and are promoted to NLO by adding the baryon SU(3) 
decuplet to the ground-state octet~\cite{Petsch17}. For $Y=\Xi$, $\Xi N_1N_2$ 
interactions occur by exciting the octet $\Xi$ hyperon to a decuplet 
$\Xi^{\ast}(1530)$ hyperon on nucleon $N_1$, $\Xi N_1\to \Xi^{\ast} N_1$, 
and deexciting it back to $\Xi$ on nucleon $N_2$, $\Xi^{\ast} N_2\to \Xi N_2$. 

The nuclear densities $\rho(r)=\rho_p(r)+\rho_n(r)$ used by us to construct 
$V_{\rm opt}^{\Xi}(\rho)$ are harmonic-oscillator type densities with the same 
radial parameters for neutrons and protons~\cite{Elton61}. The corresponding 
r.m.s. radii follow closely values derived from experiment by relating proton 
densities $\rho_p(r)$ to charge densities and including the proton charge 
finite size and recoil effects. This approach is equivalent to assigning some 
finite range to the $\Xi N$ interaction. Folding reasonably chosen $\Xi N$ 
interaction ranges other than corresponding to the proton charge radius, 
varying the spatial form of the charge density, or introducing realistic 
differences between neutron and proton r.m.s. radii, made little difference. 
In $\Xi^-_{1p}(^{12}$C), for example, all such calculated binding 
energies varied within 20\% of the $\pm$0.15~MeV measurement uncertainty 
of $B_{\Xi^-}^{1p}(^{12}$C). Finally, a fairly small absorptivity, 
Im$\,b_0=0.01$~fm in Eq.~(\ref{eq:b0}), was assumed in all of 
the fits shown and discussed below. Using only $V_{\Xi}^{(2)}$, 
$B_{\Xi^-}^{1p}(^{12}$C)=0.82$\pm$0.15~MeV was fitted by 
Re$\,b_0=0.495\pm 0.030$~fm; doubling Im$\,b_0$ was found to 
increase the fitted value of Re$\,b_0$ by only 1\%~\cite{FG21}. 
Such small absorptivities reflect the small statistical weight, 
1/16 in nuclear matter, of the $S=I=0$ $\Xi N$ channel in which 
$\Xi N\to \Lambda\Lambda$ conversion occurs.

\section{$\Xi^-$ nuclear capture events input} 
\label{sec:capture} 

\begin{figure}[!ht] 
\begin{center} 
\includegraphics[width=0.95\textwidth]{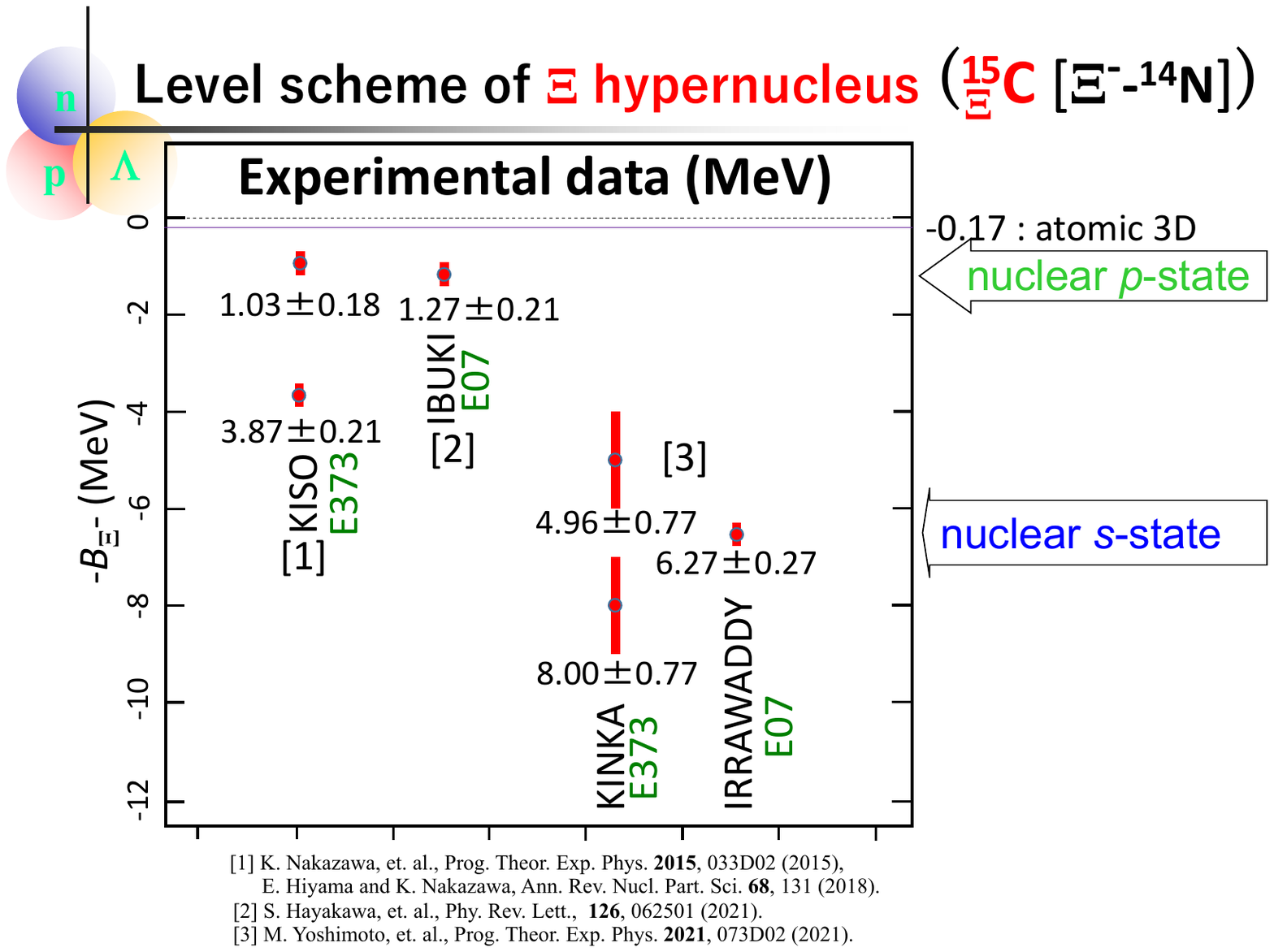} 
\caption{$\Xi^- -{^{14}{\rm N}}$ spectrum deduced from $\Xi^-$ capture events 
identified by their twin-$\Lambda$ hypernuclear decays in KEK-E373 and J-PARC 
E07 emulsion experiments, see text. Figure provided by Dr. Kazuma Nakazawa, 
based on recent results from Refs.~\cite{KISO,IBUKI,E07}.} 
\label{fig:14N} 
\end{center} 
\end{figure} 

$\Xi^-$ capture events identified during the last three decades in light 
(C,N,O) emulsion nuclei at KEK and J-PARC provide evidence for $\Xi^-$ 
nuclear states bound by about 1~MeV in $^{12}$C~\cite{E176} and in 
$^{14}$N~\cite{E176,KISO,IBUKI}. $\Xi^-$ atomic cascade in these light 
emulsion nuclei terminates normally with a $3D\to 2P$ radiative $E1$ atomic 
transition~\cite{BFG99} and the formed $\Xi^-_{1p}$ Coulomb-assisted 
nuclear states undergo hadronic capture driven by a $\Xi^-p\to\Lambda\Lambda$ 
strong-interaction conversion. Forming radiatively $\Xi^-_{1s}$ nuclear bound 
states is suppressed by almost two orders of magnitude~\cite{Zhu,Koike}. 
Hence, recent $\Xi^-_{1s}$ assignments of two capture events in $^{14}$N by 
J-PARC E07~\cite{E07} are questionable. The reported $\Xi^--{^{14}{\rm N}}$ 
spectrum is shown in Fig.~\ref{fig:14N}. Here we focus on two out of the four 
capture events, IBUKI for $\Xi^-_{1p}$ and IRRAWADDY assigned as $\Xi^-_{1s}$. 
The other two events agree with the former ones, KISO with IBUKI and KINKA 
with IRRAWADDY, provided the smaller $B_{\Xi^-}$ value is chosen for KISO and 
for KINKA respectively. 

Thus, there are two $\Xi^-_{1s}$ nuclear candidates: (i) the J-PARC E07 
$^{14}$N capture event IRRAWADDY and (ii) the J-PARC E05 hint of a bound 
state in the $^{12}$C($K^-,K^+$) inclusive spectrum, with binding energies 
\begin{equation} 
B_{\Xi^-}^{1s}({^{14}{\rm N}})=6.27\pm 0.27~{\rm MeV},\,\,\,\,\,\,\, 
B_{\Xi^-}^{1s}({^{11}{\rm B}})=8.9\pm 3.7~{\rm MeV},
\label{eq:B1s} 
\end{equation} 
where statistical and systematic uncertainties were added quadratically for 
$B_{\Xi^-}^{1s}({^{11}{\rm B}})$. We note that the more bound of these two 
$\Xi^-_{1s}$-assumed states appears to be in the lighter $^{11}$B nucleus 
rather than in the heavier $^{14}$N nucleus. This contrasts the steady 
increase with $A$ of $B_{\Lambda}(A)$ in $\Lambda$ hypernuclei~\cite{GHM16}. 
Nevertheless, as demonstrated in the next section, the fairly large 
$B_{\Xi^-}^{1s}({^{11}{\rm B}})$ uncertainty of $\pm$3.7~MeV prevents 
one from making definite statements on this issue. An improved resolution 
$(K^-,K^+)$ experiment such as J-PARC E70~\cite{Gogami24} is called for, 
as well as state-of-the-art few-body calculations of these $\Xi^-_{1s}$ 
candidates. Unfortunately, calculations using the HAL-QCD $\Xi N$ 
interaction~\cite{Hiyama20,Hiyama22} or $\chi$EFT at NLO~\cite{LHMN21} have 
not reached to date sufficiently high values of $A$ required to explore this 
issue. 

In addition to $\Xi^-_{1s}$ one needs also some $\Xi^-_{1p}$ binding-energy 
input to assess the relative weights of $V_{\Xi}^{(2)}(\rho)$, 
Eqs.~(\ref{eq:V2},\ref{eq:b0}), and $V_{\Xi}^{(3)}(\rho)$, Eq.~(\ref{eq:V3}). 
The only available $\Xi^-_{1p}$ nuclear candidates are in $^{12}$C~\cite{E176} 
and in $^{14}$N~\cite{E176,KISO,IBUKI}. However, $\Xi^-_{1p}-{^{14}{\rm N}}$ 
states are expected to spread out over a range somewhat larger than 1.2~MeV 
owing mostly to a quadrupole-quadrupole ($Q_{\Xi}\cdot Q_N$) residual 
interaction unaccounted for by $V_{\rm opt}^{\Xi}(\rho)$~\cite{FG21}. 
This $Q_{\Xi}\cdot Q_N$ interaction affects $\Xi^-_{1p}$ states attached 
to nuclear levels with total angular momentum values $J_A\geq 1$, as in 
$^{14}$N with $J^{\pi}_{\rm g.s.}=1^+$.{\footnote{Fig.~A in Ref.~\cite{FG23a} 
shows a calculated $\Xi^-_{1p}-{^{14}{\rm N}_{\rm g.s.}}$ spectrum of states 
where the IBUKI J-PARC E07 event in Fig.~\ref{fig:14N} here, bound 
by 1.27$\pm$0.21~MeV, is reproduced by a $J^{\pi}={\frac{1}{2}}^-$ 
state bound by 1.12~MeV.}} For comparison, the splitting of the 
$J^{\pi} = ({\frac{1}{2}}^-,{\frac{3}{2}}^-)$ spin-orbit doublet in 
$\Xi^-_{1p}-{^{12}{\rm C}}$ is not expected to go beyond its experimental 
uncertainty of $\pm$0.15~MeV. Hence, out of the two $\Xi^-_{1p}$ candidates 
we keep only the $\Xi^-_{1p}-{^{12}{\rm C}}$ state with $\Xi^-$ binding energy 
given by 
\begin{equation} 
B_{\Xi^-}^{1p}({^{12}{\rm C}})=0.82\pm 0.15~{\rm MeV}. 
\label{eq:B1p} 
\end{equation}

\section{Results} 
\label{sec:calc}

In this section, using $V_{\rm opt}^{\Xi}(\rho)$, we study to what extent 
the three $B_{\Xi^-}$ values listed in Eqs.~(\ref{eq:B1s},\ref{eq:B1p}) are 
compatible with each other. We also estimate spin and isospin dependent 
contributions to $B_{\Xi^-}$ of the $J^\pi =1^-$ level of 
$\Xi^-_{1s}-{^{11}{\rm B}}$ observed by J-PARC E05. Preliminary partial 
results were presented in MESON 2023~\cite{meson23}.  

\subsection{Compatibility} 
\label{subsec:comp} 

\begin{table}[!h] 
\centering
\caption{Input (boldface) and calculated values of $\Xi^-$ binding energies 
(MeV) in optical potential fits limited to $V_{\Xi}^{(2)}(\rho)$ with 
a strength parameter $b_0$ (fm). Resulting potential depths are 
$D^{(2)}_{\Xi}=-V_{\Xi}^{(2)}(\rho_0)$ (MeV) at $\rho_0=0.17\,$fm$^{-3}$.} 
\begin{tabular}{ccccc}
\hline
$B_{\Xi^-}^{1p}(^{12}$C) & $B_{\Xi^-}^{1s}(^{14}$N) 
& $B_{\Xi^-}^{1s}(^{11}$B) & $b_0$ & $D^{(2)}_{\Xi}$ \\ 
\hline 
$\bf{0.82\pm 0.15}$ & $11.8\pm 0.6$ & 9.03$\pm$0.52 & 0.495$\pm$0.030 & 
21.2$\pm$0.7 \\ 
$0.320\pm 0.004$ & $\bf{6.27\pm 0.27}$ & 4.20$\pm$0.25 & 0.247$\pm$0.011 & 
13.6$\pm$0.4 \\ 
\hline
\end{tabular}
\label{tab:targets}
\end{table}

Two $\Xi$-nuclear scenarios are outlined in Table~\ref{tab:targets}. Choosing 
$B_{\Xi^-}^{1p}(^{12}{\rm C})$, Eq.~(\ref{eq:B1p}), to fit the strength 
$b_0$ of the attractive $\Xi N$-induced component $V_{\Xi}^{(2)}(\rho)$, 
Eqs.~(\ref{eq:V2},\ref{eq:b0}), results in $\Xi^-$ nuclear binding energies 
listed in the first row. Choosing instead $B_{\Xi^-}^{1s}(^{14}{\rm N})$ 
from Eq.~(\ref{eq:B1s}) results in values listed in the second row. Clearly, 
a single $V_{\Xi}^{(2)}(\rho)$ component of $V_{\rm opt}^{\Xi}(\rho)$ is 
incapable of reproducing {\it both} given $B_{\Xi^-}^{1p}(^{12}{\rm C})$ and 
$B_{\Xi^-}^{1s}(^{14}{\rm N})$ values. Moreover, each of these two choices 
predicts completely different values for $B_{\Xi^-}^{1s}(^{11}{\rm B})$ and 
for the $\Xi$-nuclear attractive potential depth $D^{(2)}_{\Xi}$. We note in 
passing that the $\Xi^-_{1p}$--$^{12}$C capture event constrained by IRRAWADDY 
(second row) comes out hardly bound, by about 37$\pm$4~keV, with respect 
to a purely Coulomb $2P$ atomic state from which it evolves when 
$V_{\rm opt}^{\Xi}(\rho)$ is switched on. The listed calculated 
$B_{\Xi^-}^{1p}(^{12}{\rm C})$ value strongly disagrees, by about 0.5~MeV, 
with the KEK-E176~\cite{E176} input value listed in boldface in the first row. 

Complementary to Table~\ref{tab:targets} are level-width values which were 
calculated for our standard choice of absorptivity in Eq.~(\ref{eq:b0}), 
Im$\,b_0$=0.01~fm. These are $\Gamma_{\Xi^-}^{1p}(^{12}$C)=0.23~MeV and 
$\Gamma_{\Xi^-}^{1s}(^{11}$B)=0.92~MeV for line 1 in the table, and 
$\Gamma_{\Xi^-}^{1s}(^{14}$N)=0.87~MeV for line 2. These width values vary 
linearly to a good approximation with the input value of Im$\,b_0$ which was 
varied up to Im$\,b_0$=0.04~fm. The effect of absorptivity within this range 
of Im$\,b_0$ values on the calculated $B_{\Xi^-}$ values is small, well within 
the quoted experimental uncertainties. 

\begin{table}[!h]
\centering
\caption{Optical potential fits of three $B_{\Xi^-}$ input values: 
$B_{\Xi^-}^{1p}(^{12}{\rm C})$, $B_{\Xi^-}^{1s}(^{14}{\rm N})$ (boldfaced 
in Table~\ref{tab:targets}) and $B_{\Xi^-}^{1s}(^{11}{\rm B})=8.9\pm\delta B$ 
(MeV). $\chi^2$ values are listed along with fitted strength parameters 
$b_0,B_0$ (fm) and their respective $\Xi$-nuclear potential depths: 
$D_{\Xi}^{(2)}=-V_{\Xi}^{(2)}(\rho_0)$, $D_{\Xi}^{(3)}=V_{\Xi}^{(3)}(\rho_0)$ 
(MeV) at $\rho_0=0.17\,$fm$^{-3}$. The total depth (positive) is 
$D_{\Xi}=D_{\Xi}^{(2)}-D_{\Xi}^{(3)}$.} 
\begin{tabular}{ccccccc}
\hline 
$\delta B$ & $\chi^2$ & $b_0$ & $B_0$ & $D_{\Xi}^{(2)}$ & $D_{\Xi}^{(3)}$ & 
$D_{\Xi}$ \\ 
\hline 
3.7 & 0.99 & 1.13$\pm$0.20 & 0.54$\pm$0.08 & $31.0^{+1.8}_{-2.2}$ & 
29.0$\pm$4.3 & $2.0^{+2.1}_{-2.5}$ \\ 
1.4 & 6.80 & 1.16$\pm$0.51 & 0.54$\pm$0.21 & $31.3^{+3.8}_{-6.8}$ & 
29.0$\pm$11.3 & $2.3^{+4.5}_{-7.5}$ \\ 
\hline 
\end{tabular} 
\label{tab:11B} 
\end{table} 

Next, we add $B_{\Xi^-}^{1s}(^{11}{\rm B})=8.9\pm\delta B$ (MeV) to the input 
values $B_{\Xi^-}^{1p}(^{12}{\rm C})$ and $B_{\Xi^-}^{1s}(^{14}{\rm N})$ 
boldfaced in Table~\ref{tab:targets}, attempting to fit these three binding 
energies simultaneously in terms of two strength parameters, $b_0$ and 
$B_0$, of an attractive $\Xi N$-induced component $V_{\Xi}^{(2)}(\rho)$ plus 
a repulsive $\Xi NN$-induced component $V_{\Xi}^{(3)}(\rho)$, respectively, 
of $V_{\rm opt}^{\Xi}$. This gives rise to a substantial repulsive partial 
depth $D_{\Xi}^{(3)}$ that cancels out almost completely the excessively large 
attractive partial depth $D_{\Xi}^{(2)}$, as shown in Table~\ref{tab:11B} 
for two choices of the experimental uncertainty: $\delta B$=3.7~MeV for the 
J-PARC E05 nominal uncertainty, and reducing it to 1.4~MeV by suppressing 
its systematic component. We note that the fitted $b_0$ and $B_0$ are 100\% 
anti-correlated and hence the partial potential depths $D_{\Xi}^{(2)}$ and 
$D_{\Xi}^{(3)}$ are also 100\% anti-correlated. 

The $\chi^2\approx 1$ listed in the first row of Table~\ref{tab:11B} is 
perfectly acceptable for a two-parameter three-data-points fit. The relatively 
large fitted values for both $b_0$ and $B_0$ are needed to make the input 
values of $B_{\Xi^-}^{1p}(^{12}{\rm C})$ and $B_{\Xi^-}^{1s}(^{14}{\rm N})$ 
compatible with each other. The high value of $\chi^2\approx 7$ for one degree 
of freedom in the second-row fit makes it unacceptable, implying that an 
improved resolution $(K^-,K^+)$ experiment is necessary to demonstrate the 
incompatibility of the present values of $B_{\Xi^-}^{1s}(^{11}{\rm B})$ and 
$B_{\Xi^-}^{1s}(^{14}{\rm N})$. 

Finally, we note that the $\Xi NN$-induced repulsive $V_{\Xi}^{(3)}$ 
contribution to the $B_{\Xi^-}^{1s}$ values listed in Table~\ref{tab:11B} 
is huge, about 13$\pm$1~MeV in $^{11}$B and $^{14}$N, as estimated by 
switching off $V_{\Xi}^{(3)}$ in the $\Xi^-$ bound-state calculations. 
This is roughly factor of 3 larger than $\Lambda NN$ repulsive contributions 
evaluated long time ago in $p$-shell $\Lambda$ hypernuclei~\cite{GSD}, 
consistently with the corresponding ratio of three-body repulsive potential 
depths: $D_{\Xi}^{(3)}\sim 30$~MeV listed in the table to $D_{\Lambda}^{(3)}
\sim 10$~MeV repulsion obtained by fitting recently $B_{\Lambda}$ data across 
the periodic table~\cite{FG23b,FG23c}. A ratio that large goes against one's 
normal expectation that $YN$ and $YNN$ potential contributions decrease with 
increasing strangeness $|{\cal S}|$~\cite{DF92}. 

\subsection{Spin and isospin contributions to $B_{\Xi^-}^{1s}(^{11}{\rm B})$} 
\label{subsec:SU(4)} 

To estimate spin and isospin contributions to $B_{\Xi^-}^{1s}(^{11}{\rm B})$ 
we start from the most general spin and isospin dependent $s$-wave $\Xi N$ 
interaction: 
\begin{equation}
V_{\Xi N}=V_0+V_{\sigma}{\vec\sigma}_{\Xi}\cdot{\vec\sigma}_N+V_{\tau}
{\vec\tau}_{\Xi}\cdot{\vec\tau}_N+V_{\sigma\tau}{\vec\sigma}_{\Xi}\cdot
{\vec\sigma}_N\,{\vec\tau}_{\Xi}\cdot{\vec\tau}_N \, .
\label{eq:V2body}
\end{equation}
The $\Xi$ nucleus optical potential $V_{\Xi}^{(2)}(\rho)$ used in 
Table~\ref{tab:targets} to evaluate values of $B_{\Xi^-}^{1s}(^{11}{\rm B})$ 
accounts only for the spin-isospin independent term $V_0$ of $V_{\Xi N}$ in 
Eq.~(\ref{eq:V2body}). It disregards spin-isospin dependent terms specified by 
$V_{\sigma}$, $V_{\tau}$ and $V_{\sigma\tau}$. Using HAL-QCD~\cite{HALQCD20} 
ratios of volume integrals of $V_{\sigma}(r_{\Xi N})$, $V_{\tau}(r_{\Xi N})$ 
and $V_{\sigma\tau}(r_{\Xi N})$ to that of $V_0(r_{\Xi N})$, we estimate 
energy shifts and splitting of the $1s_{\Xi^-}$ $J^\pi=(1^-,2^-)$ g.s. doublet 
levels in $\Xi^--{^{11}{\rm B}}$. We find that the $V_{\tau}$ term makes 
both $1^-$ and $2^-$ levels less bound by about 0.5~MeV with respect to 
the $B_{\Xi^-}^{1s}(^{11}{\rm B})$ value listed in Table~\ref{tab:targets}, 
whereas the $V_{\sigma}$ and $V_{\sigma\tau}$ terms push the $1^-$ level 
about 0.5~MeV deeper, back to $B_{\Xi^-}^{1s}(^{11}{\rm B})=9$~MeV listed 
in Table~\ref{tab:targets}. The $2^-$ level becomes then less bound by about 
0.3~MeV, lying altogether about 0.8 MeV above the $\Xi^-_{1s}-{^{11}{\rm B}}$ 
$1^-$ g.s.

\section{Concluding remarks} 
\label{sec:concl} 

In this work we explored to what extent a $\Xi^-_{1s}$ assignment for the 
recent `hint' of a $\Xi^- - {^{11}{\rm B}}$ nuclear state in the J-PARC E05 
${^{12}}{\rm C}(K^-,K^+){^{12}_{\Xi}}{\rm Be}$ end-point spectrum~\cite{E05} 
is compatible with $\Xi^-_{1s}$ and $\Xi^-_{1p}$ nuclear-state 
interpretations of $\Xi^-$ capture events seen in light 
emulsion-nuclei~\cite{E176,KISO,IBUKI,E07}. We used for this purpose 
a density-dependent optical-potential methodology applied by us recently 
to $\Lambda$ hypernuclear states extracted from several types of production 
experiments across the periodic table~\cite{FG23b,FG23c}. This methodology was 
applied as well to $\Xi^-$ hypernuclei using scarce $\Xi^-$ capture events 
data derived by observing weak-decay products following $\Xi^-$ capture in 
light emulsion nuclei~\cite{FG21,FG23a}. Unfortunately, assignments made to 
such $\Xi^-$-nuclear states are not necessarily unique. 

One conclusion that may be drawn from the present work is that confirming 
a $\Xi^-$ nuclear state with $B_{\Xi^-}\sim 9$~MeV in a good-resolution 
${^{12}}{\rm C}(K^-,K^+)$ experiment will rule out definitively 
a $\Xi^-_{1s}-{^{14}{\rm N}}$ assignment for the KINKA and IRRAWADDY 
$\Xi^-$ capture events in $^{14}$N. A perfectly natural assignment of 
$\Xi^0_{1p}-{^{14}{\rm C}}$ for these events was already discussed by 
us in Ref.~\cite{FG23a}. Another consequence of confirming 
$B_{\Xi^-}^{1s}(^{11}{\rm B})\sim 9$~MeV would be that only little $\Xi NN$ 
three-body contribution is needed, if at all, to determine the $\Xi$-nuclear 
potential depth at normal nuclear density, here found to be approximately 
21~MeV (attraction). 

Our well-depth value is within the range of values 17$\pm$6~MeV derived in 
Ref.~\cite{HH21} from a $K^+$ quasi-free spectrum measured by BNL-AGS E906 
in ($K^-,K^+$) on $^9$Be~\cite{E906}, covering also deductions made from 
KEK-E224~\cite{E224} and BNL-AGS E885~\cite{E885} on $^{12}$C targets. 
On the other hand, a value of 21~MeV is considerably larger than reported 
in some recent model evaluations, e.g., HAL-QCD~\cite{Inoue19}: 4~MeV, 
$\chi$EFT$@$NLO~\cite{Kohno19}: 9~MeV, and SU(6)~\cite{GC21}: 6~MeV.  
A notable exception is provided by versions ESC16*(A,B) of the latest 
Nijmegen extended-soft-core $\Xi N$ interaction model~\cite{Nij20} where 
$\Xi$-nuclear potential depths larger than 20~MeV are derived. However, these 
large values are reduced substantially by $\Xi NN$ three-body contributions 
within the same ESC16* model.

\section*{Acknowledgments}

This work started as part of a project funded by the European Union's 
Horizon 2020 research \& innovation programme, grant agreement 824093.

\end{document}